\documentclass
[preprint,prl,showpacs,onecolumn,9pt,superscriptaddress]{revtex4}%
\usepackage{amsfonts}
\usepackage{amsmath}
\usepackage{amssymb}
\usepackage{graphicx}%
\begin{document}
\title{Quantum mechanical implementation of DNA algorithm for satisfiability problem}
\author{T. T. Ren}
\affiliation{State Key Laboratory of Magnetic Resonance and Atomic and Molecular Physics,
Wuhan Institute of Physics and Mathematics, Chinese Academy of Sciences, Wuhan
430071, China}
\affiliation{Graduate School of the Chinese Academy of Sciences, Beijing 100049, China}
\author{M. Feng}
\email{mangfeng@wipm.ac.cn}
\affiliation{State Key Laboratory of Magnetic Resonance and Atomic and Molecular Physics,
Wuhan Institute of Physics and Mathematics, Chinese Academy of Sciences, Wuhan
430071, China}
\author{W.-L. Chang}
\email{changwl@cc.kuas.edu.tw}
\affiliation{Department of Computer Science and Information Engineering, National Kaohsiung
University of Applied Sciences, Kaohsiung City, Taiwan 807-78, China}
\author{J. Luo }
\affiliation{State Key Laboratory of Magnetic Resonance and Atomic and Molecular Physics,
Wuhan Institute of Physics and Mathematics, Chinese Academy of Sciences, Wuhan
430071, China}
\author{M. S. Zhan}
\affiliation{State Key Laboratory of Magnetic Resonance and Atomic and Molecular Physics,
Wuhan Institute of Physics and Mathematics, Chinese Academy of Sciences, Wuhan
430071, China}

\begin{abstract}
DNA computation could in principle solve the satisfiability (SAT) problem due
to the operations in parallel on extremely large numbers of strands. We
demonstrate some quantum gates corresponding to the DNA ones, based on which
an implementation of DNA algorithm for SAT problem is available by quantum
mechanical way. Since quantum computation owns the favorable feature of
operations in parallel on $2^{n}$ states by using only $n$ qubits, instead of
$2^{n}$ strands in DNA computation, computational complexity is much reduced
in treating the SAT problem quantum mechanically. We take a three-clause SAT
problem with two variables as an example, and carry out a NMR experiment for
solving a one-variable SAT problem.

\end{abstract}

\pacs{03.67.Ac, 89.70.Eg, 87.14.gk, 87.10.Tf}
\maketitle

Recent years have witnessed some outstanding breakthroughs in the molecular
computation proposed by Feynman \cite{Feyn1} in 1961. For example, one of the
famous non-deterministic polynomial (NP)\ problems, i.e., satisfiability (SAT)
problem, has been in principle worked out by biological computation with DNA
strands \cite{Adle,Lipton}. As it is generally considered that other NP
problems could be reduced to a solvable SAT problem \cite{Cook}, the achieved
solutions by the DNA-based biological computation (DNAC) present us hopes to
solve all the NP problems.

On the other hand, quantum computation (QC), another proposal by Feynman
\cite{Feyn2}, has drawn much attention over past decades. Eight atomic qubits
\cite{Blatt} and six photonic qubits \cite{Pan} have been entangled so far,
respectively, and simple quantum algorithms have been tested experimentally
\cite{Blatt1}. It is believed that QC outperforms classical computation in
treating some NP problems \cite{Deutsch,Shor,Grover,Hogg,Zhu}.

DNAC could potentially have vastly more parallelism than conventional
classical computations, which makes it possible to solve the SAT problem in
principle. In contrast, QC, running in an intrinsically different mechanism,
works on qubits which would\ be encoded in states $|0\rangle$ and $|1\rangle$
as well as in arbitrary superposition of $|0\rangle$ and $|1\rangle.$ As a
result, the computation in parallelism in QC could be done naturally by
superposition of states in a single system. Therefore, to represent $2^{n}$
states, we need $2^{n}$ DNA strands in computation, but only $n$ qubits
quantum mechanically. Besides, entanglement is the unique feature in QC, which
is the base of quantum logic gates and related to nonlocality. Another feature
of QC, different from DNAC, is the state collapse due to measurement. To keep
a qubit unchanged after measurement, we have to employ auxiliary qubits.

The present work focuses on finding some relations between DNAC and QC, based
on which the solution of the SAT problem by DNAC algorithm would be carried
out quantum mechanically. As a parallel implementing computation, DNAC could
in principle solve a SAT problem with extremely large number of strands (i.e.,
bits). In contrast, the QC, running on much less resource of qubits, should be
able to solve the same problem much more efficiently even following the same
computing route. So once we could find the correspondence between the basic
operations of DNAC and QC, a translation of the DNA algorithm to quantum
version will make us available to try a quantum mechanical implementation of
DNA algorithm. We argue that it would help us find new functions of QC and new
ways to quantum algorithm even if such a DNAC-based quantum mechanical
implementation would not really reduce NP problems to P problems. On the other
hand, DNAC would be further understood from our study with QC. We will also
test our quantum treatment by Nuclear Magnetic Resonance (NMR) experiment.

We first review briefly the basic operations in DNAC \cite{Guo}: $Append$,
$Extract$, $Discard$, $Amplify,Merge,Detect$ and $Read$. The operation
$Append$, including $Append-Head$ and $Append-Tail$, is to put a short DNA
strand to the head and the tail of a long strand, respectively. That is to
say, $Append-Head(B,u_{j})=\{u_{j},B_{n},B_{n-1},...B_{2},B_{1}\},$ and
$Append-Tail(B,u_{j})=\{B_{n},B_{n-1},...B_{2},B_{1},u_{j}\},$ with $B$ a set
consisting of a number of elements $B_{k}$ $(k=1,....,n).$ $Extract$ is to
extract some of the required DNA strands. In most operations, $Extract$
results in a separation of one tube into two with one tube involving the
required strands and the other involving the rest. The corresponding formulas
are $+\{U,u_{j}^{1}\}=\{u_{n},u_{n-1},...,u_{j}^{1},...,u_{2},u_{1}\}$ and
$-\{U,u_{j}^{1}\}=\{u_{n},u_{n-1},...,u_{j}^{0},...,u_{2},u_{1}\}$ with $U$
the set involving elements $u_{k}$ $(k=1,....,n)$ and $u_{j}^{1}$ and
$u_{j}^{0}$ denoting values of $u_{j}$ to be 1 and 0, respectively. $Discard$
is to null a tube, i.e., removing each DNA strand from the tube. $Amplify$
replicates all of the DNA strands in the test tube, which creates a number of
identical copies and then $Discard$ the original one. $Merge$ corresponds to
the operation to pour many tubes of DNA strands into one tube without any
change in the individual strands, which could be described by $\cup
(P_{1},P_{2},...P_{n})=P_{1}\cup P_{2}\cup...\cup P_{n},$ with $P_{k}$
$(k=1,2,...,n)$ being a tube with DNA stands$.$ $Detect$ leads to a result
'YES' once there is at least one DNA strand in the tube, or 'NO' otherwise.
$Read$ gives an explicit description of one DNA strand no matter how many
molecules in the tube.

\bigskip On the side of QC, there are some basic operations constituting
universal QC \cite{Nielson}, where the most frequently mentioned gates are
$\mathbf{R}(\theta)=\left(
\begin{array}
[c]{cc}%
1 & 0\\
0 & e^{i\theta}%
\end{array}
\right)  $ for the qubit encoding $|0\rangle=\binom{1}{0}$ and $|1\rangle
=\binom{0}{1},$ Hadamard gate $\mathbf{H}$=$\left(
\begin{array}
[c]{cc}%
1 & 1\\
1 & -1
\end{array}
\right)  /\sqrt{2}$ to change $|0\rangle$ to $(|0\rangle+|1\rangle)/\sqrt{2}$
and $|1\rangle$ to $(|0\rangle-|1\rangle)/\sqrt{2},$and controlled-NOT gate
$\mathbf{CNOT}=\left(
\begin{array}
[c]{cccc}%
1 & 0 & 0 & 0\\
0 & 1 & 0 & 0\\
0 & 0 & 0 & 1\\
0 & 0 & 1 & 0
\end{array}
\right)  $. To be more efficient, we sometimes employ three-qubit Toffoli gate
$\mathbf{TOFF}$ to flip the target qubit when the two control qubits are both
in states $|1\rangle.$

Comparing QC with DNAC, we could find some relations between them. Quantum
mechanically, $Append$ could be described as a tensor product, i.e.,
$Append-Head(B,u_{j})=\{u_{j}\}\otimes\{B\}$ and $Append-Tail(B,u_{j}%
)=\{B\}\otimes\{u_{j}\}.$ The operation $Extract$ could, to some extent, be
carried out by $\mathbf{CNOT}.$ On the other hand, a Hadamard gate in QC could
be carried out by the operations of DNAC with $Extract$ to separate two
subsets respectively including $|0\rangle$ and $|1\rangle,$and then with
$Append$ and $Merge$ to realize $|0\rangle\rightarrow(|0\rangle+|1\rangle
)/\sqrt{2}$ and $|1\rangle$ $\rightarrow$ $(|0\rangle-|1\rangle)/\sqrt{2}.$ To
be specific, we give an example below to simulate quantum superposition by
operations in DNAC. We initially have an empty set \{$\phi$\}, and replicate
it by $Amplify\{\phi\}$ to be two empty sets. $Append-Tail\{\phi,|0\rangle\}$
and $Append-Tail\{\phi,|1\rangle\}$ yield the sets $\{|0\rangle\}$ and
$\{|1\rangle\}$, respectively. After the operation $Merge,$ we could have a
superposition in the set $\{|0\rangle+|1\rangle\},$ equivelent to
$\mathbf{H}|0\rangle$ in QC$.$ Repeating above steps, we could also get the
set $\{(|0\rangle+|1\rangle)|0\rangle+(|0\rangle+|1\rangle)|1\rangle\},$
actually corresponding to $\mathbf{H}|0\rangle\mathbf{\otimes H}|0\rangle$ in
QC$.$ Nevertheless, it seems that DNAC could not fully accomplish the jobs by
QC. For example, the QC operation $\mathbf{R}(\theta)$ with $0<\theta<2\pi$
could not be efficiently simulated by DNAC. But QC could carry out any job by
DNAC in a more efficient way.

In what follows, we will solve a SAT problem quantum mechanically following
the route in DNAC. As mentioned above, QC using qubits could save the resource
from $2^{n}$ DNA strands to $n$ qubits. Even if there are auxiliary qubits
involved, the number of qubits increases only linearly with the size of the QC
task. Let's consider a simple case as an example with the formula%
\begin{equation}
F=(u_{2}\vee u_{1})\wedge(\overline{u_{2}}\vee\overline{u_{1}})\wedge(u_{1}),
\end{equation}
where $u_{2\text{ }}$and $u_{1\text{ }}$are Boolean variables whose values can
be $0$ (false) or $1$ (True). $\vee$ is the \textquotedblleft logical
OR\textquotedblright\ operation with $u_{2}\vee u_{1}=0$ only if $u_{2}%
=u_{1}=0,$ and $\wedge$ is the \textquotedblleft logical AND\textquotedblright%
\ operation with $u_{2}\wedge u_{1}=1$ only if $u_{2}=u_{1}=1$. $\overline
{u_{2}}$ and $\overline{u_{1}}$ are the operations \textquotedblleft
NEGATION\textquotedblright\ of $u_{2}$ and $u_{1}$, respectively, i.e.,
$\overline{u_{2}}$ being 0 if $u_{2}=1$ and being 1 if $u_{2}=0$. The
satisfiability problem is to find appropriate values for $u_{2}$ and $u_{1}$
to make the formula F true.

The logic AND and OR could be carried out by quantum circuits \cite{Vedral},
as shown in Fig. 1. To solve Eq. (1), we employ four quantum registers
$|u_{2}u_{1}\rangle$, $|y_{2}y_{1}\rangle$, $|r_{2}r_{1}r_{0}\rangle$ and
$|c_{3}c_{2}c_{1}c_{0}\rangle$, which are introduced basically from the idea
of DNAC \cite{Lipton}. $u_{2}$ and $u_{1}$ are qubits initially zero and then
in superposition by Hadamard gates. $y_{2}$ and $y_{1},$ also initially being
zero, are auxiliary qubits acting as copies of $u_{2}$ and $u_{1},$
respectively. The third register stores the results of OR, where the qubits
inside are initially prepared to be $r_{2}^{1},$ $r_{1}^{1},$ and $r_{0}^{0},$
with the superscripts being values of the qubits. After each time with the
data transferred to the third register from the second one, $y_{2}$ and
$y_{1}$ will be nulled for later use. The fourth register, including four
qubits, are used for storing the results of AND, where except
$\vert$%
$c_{0}^{1}\rangle,$ other qubits are initially zero. After each AND operation,
the results are stored in $|c_{3}\rangle,$ $|c_{2}\rangle,$ or $|c_{1}%
\rangle,$ respectively, and we have to restore the qubits in the third
register to be $r_{2}^{1},$ $r_{1}^{1},$ and $r_{0}^{0}$ for later use
repeatedly. With these ideas, to accomplish an evaluation of Eq. (1), we
design a quantum circuit in Fig. 2, where the qubits are input from the
left-hand side of the circuit. We get started from the input state $|u_{2}%
^{0}u_{1}^{0}\rangle|y_{2}^{0}y_{1}^{0}\rangle|r_{2}^{1}r_{1}^{1}r_{0}%
^{0}\rangle|c_{3}^{0}c_{2}^{0}c_{1}^{0}c_{0}^{1}\rangle.$ Following the gates
in Fig. 2 step by step, we finally obtain, after a measurement on $|c_{3}%
^{1}\rangle,$ $|u_{2}^{0}u_{1}^{1}\rangle|y_{2}^{0}y_{1}^{0}\rangle|r_{2}%
^{1}r_{1}^{1}r_{0}^{0}\rangle|c_{2}^{1}c_{1}^{1}c_{0}^{1}\rangle,$ implying
the correct values of $u_{2}$ and $u_{1}$ to be 0 and 1, respectively.

We will below employ NMR approach to check our theory experimentally. Although
the quantum information processed by NMR is made on the ensemble of nuclear
spins, instead of individual spins, NMR has remained to be the most convenient
experimental tool to demonstrate quantum information processing due to its
mature and well-controllable technology \cite{Chuang}. We will employ spatial
averaging method \cite{Jones} to prepare the thermal equilibrium ensemble to
the pseudo-pure state. To make the experimental operations simple and
reliable, we will carry out below a three-qubit case corresponding to a
solution of the simplest SAT problem $F=(u_{1}),$ i.e., a SAT with one clause
involving only a single variable. The quantum circuit is plotted in Fig. 3,
where $|u_{1}\rangle$ is the qubit holding the variable, $|y_{1}\rangle$ is
the copy of $|u_{1}\rangle,$ and $|c_{1}\rangle$ is to store the evaluating
result. Following the steps in Fig. 3, we could obtain the output
$(|000\rangle+|101\rangle)/\sqrt{2}.$ By a measurement on $|c_{1}^{1}\rangle,$
we could obtain $|u_{1}^{1}\rangle$ and $|y_{1}^{0}\rangle,$ which means that
the evaluation of $u_{1}$ should be one and $y_{1}$ has been nulled for later use.

We have carried out the quantum circuit experimentally on a Varian INOVA 500
NMR spectrometer with the sample $^{13}C-labelled$ alanine, i.e., $_{1}%
^{13}CH_{3}-_{2}^{13}CH(NH_{2})-_{3}^{13}COOH$. The three qubits are encoded
in the carbons $_{1}^{13}C,$ $_{2}^{13}C,$ $_{3}^{13}C,$\ respectively, with
$J$-coupling constants $J_{12}=34.79$ Hz$,$ $J_{23}=54.01$ Hz$,$ and
$J_{13}=1.20$ Hz. The pulse sequences to prepare the pseudo-pure state are
from \cite{Cory}. The Hadamard gate can be realized by a single $\pi/2$ pulse
along the $x$ axis and $\mathbf{CNOT}$ is implemented by the pulses
\cite{Jones} $[\pi/2]_{y}^{2}\rightarrow(1/4J)\rightarrow\lbrack\pi]_{x}%
^{1,2}\rightarrow(1/4J)\rightarrow\lbrack\pi]_{x}^{1,2}\rightarrow\lbrack
\pi/2]_{x}^{2}$. However, due to weak measurement in NMR, we have no state
collapse after a measurement. Besides, only single quantum coherence can be
detected in NMR. As a result, we have to employ some additional operations for
detecting the output state $(|000\rangle+|101\rangle)/\sqrt{2}$. We may detect
the second qubit directly by applying a $\pi/2$ readout pulse along the $x$
axis, yielding Fig. 4(b). But for the first and third qubits, we need to
disentangle them before measurement. To this end, we apply a $\mathbf{CNOT}%
$\textbf{ }gate, respectively, on the first and second qubits followed by
another $\mathbf{CNOT}$ gate, respectively, on the second and first qubits to
get the state $(|000\rangle+|011\rangle)/\sqrt{2}$. Then the first qubit can
be read out by a single $\pi/2$ pulse along the $x$ axis, as shown in Fig.
4(a). Similar steps applied to the third qubit result in the spectrum in Fig. 4(c).

The experimental results are in good agreement with our theoretical
prediction, which proves the SAT problem to be solvable by QC. Some remarks
must be addressed. First of all, the three-qubit NMR experiment we have
carried out suffices to make a comprehensive test for our theory, because we
have achieved the key aspects of our theory. Although the simple cases with
eleven and three qubits, respectively, did not reflect the efficiency of QC
implementation for SAT problem, we argue that, with more variables and clauses
involved, the QC efficiency would be more and more evident, which could also
be found in our later discussion about the computational complexity. Secondly,
DNAC does not involve entanglement, whereas entanglement does appear in our
quantum treatment. The necessity of additional operations to disentangle the
output qubits is not the intrinsic characteristic of our quantum mechanical
treatment, but due to the unique feature of NMR technique. Anyway, those
additional operations have not changed the essence of our implementation.
Thirdly, although it is workable in solving SAT problems, DNAC has been lack
of mathematical description. In this sense, our investigation of the relation
between DNAC and QC actually presents a mathematical description of the DNAC
operations, which is helpful for us to further understand the efficiency and
the functions of DNAC.

It is very difficult to discover a quantum algorithm with exponential
speed-up. That is why the frequently mentioned quantum algorithms have been
only few so far. We argue that, even if it does not provide a general way to
reduction of the NP problem to a P problem, our quantum version of the DNAC
algorithm should be able to efficiently reduce the computational complexity,
compared to the original DNAC treatment. We have simply assessed the
computational complexity of our quantum treatment from Fig. 2 and more general
consideration for the SAT problem with $m$ clauses and $n$ variables
\cite{Explain}: The time complexity is $O(n)$ $\mathbf{H}$ gates, $O(6\times
m\times n)$ $\mathbf{NOT}$ gates, $O(2\times m\times n)$ $\mathbf{CNOT}$
gates, $O(m\times n+m)$ $\mathbf{TOFF}$ gates, and $O(1)$ projective
operations for measurement. The space complexity is $O(m+3\times n+2)$ qubits,
invoving the qubits for ancillary. More strict proof in detail will be
published elsewhere.

In summary, we have demonstrated a quantum mechanical implementation of DNAC
to solve a SAT problem. Both QC and DNAC are hot topics as interdisciplinary
subjects, and both of them have merits and drawbacks \cite{Adle,Lipton,Benn}.
Our investigation has presented the relations between them, and we argue that
quantum treatment could reduce the complexity of the solution to some NP
problems. The relations we have presented between QC and DNAC could enable not
only a further exploration of new ways to QC algorithm, but also a further
understanding of DNAC from a brand-new angle.

The work is partly supported by NNSFC under Grant No. 10774163, by NFRPC under
Grant No 2006CB921203, and partly by NSC under Grants No. 96-2221-E-151-008-
and 96-2218-E-151-004-.


\begin{thebibliography}{99}
\bibitem {Explain}A strict mathematical proof is under preparation for a
general consideration of quantum mechanical solution to SAT problems in
comparison with Lipton's work \cite{Lipton}, which will be published elsewhere.
\end{thebibliography}
\end{document}